# Sub-ns spin-transfer switching: compared benefits of free layer biasing and pinned layer biasing

T. Devolder[a], C. Chappert[a], and K. Ito[b]

*Institut d'Electronique Fondamentale, CNRS UMR 8622, Bât. 220, université Paris-Sud, 91405 Orsay, France.*

*Hitachi Cambridge Laboratory, Hitachi Europe, Ltd., Cavendish Laboratory, Madingley Road, Cambridge CB3 0HE, UK*

***Abstract***: *We analyze the statistical distribution of switching durations in spin-transfer switching induced by current steps, and discuss biasing strategies to enhance the reproducibility of switching durations. We use a macrospin approximation and model the effect of finite temperature as a Boltzmann distribution of initial magnetization states (adiabatic limit). We compare three model spin-valves: a spin-valve with a free layer whose easy axis is parallel to the pinned layer magnetization (standard geometry), a pinned layer with magnetization tilted with respect to the free layer easy axis (pinned layer biasing), and a free layer whose magnetization is pulled away from easy axis by a hard axis bias (free layer biasing). In the conventional geometry, the switching durations follow a broad regular distribution, with an extended long tail comprising very long switching events. For the two biasing strategies, the switching durations follow a multiply-stepped distribution, reflecting the precessional nature of the switching, and the statistical number of precession cycles needed for reversal. We derive analytical criteria to avoid switching events lasting much longer than the average switching duration, in order to achieve the highest reproducibilities. Depending on the current amplitude and the biasing strength, the width of the switching time distribution can be substantially reduced, the best reproducibility being achieved for free layer biasing at overdrive current of a few times unity.*

## 1. Introduction

The spin-transfer effect [1] results from the exchange of angular momentum between a spin-polarized electrical current and the magnetization of a nanomagnet. The spin-transfer results in torques that can be used to manipulate magnetic configurations with a sole current, leading to new phenomena such as the displacement of domain walls [2], the generation of spin waves [3], and the pumping of high amplitude steady magnetization precessions [4].

The spin torque can also simply switch the magnetization of a uniaxial nanomagnet [5], which is considered as a promising route for memory applications, since this type of switching has proven sub-ns potential [6, 7, 8]. However, previous investigations have concluded that the reversal speed in the sub-ns regime had insufficient reproducibility. This has often been interpreted qualitatively as resulting from classical thermal fluctuations. However, a descriptive model with a transparent formalism was not yet available so far, such that routes to ameliorate the reproducibility of the switching durations could not be proposed yet.

In a previous experimental study [9], we have shown that the sub-ns pulse durations leading to successful switching events are discrete durations reflecting the precessional nature of magnetization dynamics, and the topological peculiarities in the set of possible magnetization trajectories. We had also shown that this tendency towards quantization of the switching times could be manipulated by biasing the free layer using an externally applied hard axis field to lift the near degeneracy between magnetic trajectories, and to change the switching reliability in the sub-ns regime.

In this paper, we build on our previous experimental findings, and extend a model that takes into account the precessional dynamics, the spin-torque and the main characters of thermal fluctuations within the macrospin approximation and the adiabatic limit. We model the statistical distributions of the switching duration in three different situations, two of them having being formerly experimentally characterized in our group [9]. Our reference is the standard configuration (SC) of spin-valves with their free layer easy axis parallel to the direction of the pinned layer magnetization. We also consider the Free Layer Biasing Configuration (FLBC), where the free layer is subjected to a hard axis field either applied externally as done in ref [9], or applied by an exchange bias acting on the free layer. We finally also consider the Pinned Layer Biasing Configuration (PLBC), where the pinned layer is pinned at a given angle with the free layer easy axis, in their common plane as done experimentally for instance in ref. [10]. We evaluate numerically the distributions of the switching duration in those three different configurations, and discuss the results using analytical modeling that sheds light onto the main physical origins of the switching time fluctuations. We finally determine how much biasing and how much margin should be taken in the current pulse duration to achieve error-free switching. We also report criteria to decide what the optimal biasing angle and current amplitude are.

The paper is organized as follows. Section 2 gathers the description of the systems we consider, and the equations describing their dynamical behaviors. Section 3 describes the main characters of switching in the standard configuration. Section 4 summarizes how we model the effect thermal fluctuations on the magnetic behavior. The resulting distribution of switching time in the standard configuration is calculated and discussed. Section 5 is dedicated to main characters of switching in the two biased configurations, before their associated statistical distribution of switching duration is derived in Section 6. Concluding remarks are offered at the end of the article.

## 2. Definitions and dynamical equations of magnetization motion

### 2.1. Definition of the biasing configurations

The goal of this article is to discuss the differences in switching time reproducibility in three different situations. They are illustrated in Figure 1. Our reference configuration will be quoted as the "standard configuration" (SC, Figure 1A), which consist of a spin-valve with a free layer having an easy axis along *(x)*, and a pinned layer that is magnetized also along *(x)*. The free layer is supposed not to feel any external field. Our second situation (Figure 1B) is the one studied experimentally by Krivorotov et al. [10], and will be quoted as the Pinned Layer Biasing Configuration (PLBC). In this configuration (Figure 1B), the pinned layer is pinned at a given angle with respect to the free layer easy axis. Both magnetization lie in the sample plane and the free layer is still supposed not to feel any external field. The third configuration is the one we have formerly studied in [8, 9], and will be referred as the Free Layer Biasing Configuration (FLBC, Figure 1C). In this configuration the free layer is subjected to a static hard axis field applied either externally or by an exchange bias. The pinned layer is magnetized along *(x)*, i.e. parallel to the easy axis of the free layer.

When we will make *direct* comparisons between the PLBC and the FBLC, we will always consider biasing parameter leading to the *same* tilting angle of either the pinned layer magnetization or the free layer magnetization.

Throughout this article, we will consider switching of free layer states magnetized initially near the positive easy axis ($m_x$>0) to the reversed orientation, as a result of steps of current applied infinitely abruptly at a time *t*=0. We define the switching duration as the precise instant when the hard axis is overcome, i.e. when the condition $m_x$=0 is fulfilled.

### 2.2. Dynamical equations

We model the free layer by a thin macrospin lying in the *(xy)* plane, having a dimensionless magnetization **m**=**M**/*Ms* with saturation magnetization $\mu_0 M_S$=1 T, a thickness *t*=2.8 nm, and a surface 0.02 µm² in the *(xy)* plane. The anisotropy is assumed uniaxial with an easy axis *(x)* and a strength described by an anisotropy field of $\mu_0 H_k$=10 mT. We write $h_k = H_k/M_s$. The demagnetization tensor is assumed to be diagonal, with a sole unit component along *(z)*. The Gilbert damping parameter is assumed to be α=0.02. All these material parameters are chosen near the experimental data from ref [8, 9]. In the FLBC, we apply a hard axis field $H_y=h_y M_S$ along *(y)*, keeping $\|\mathbf{H_y}\| < H_k$. We denote $\mathbf{H_{eff}}$ the sum of the anisotropy field, the demagnetizing field and the applied field. The applied current density $J_{applied}$ is along *(z)* and it is assumed to carry a spin polarisation parallel to the magnetization of the pinned layer, with spin polarisation amplitude $\Pi=0.135$. To account for the non collinearity between the free layer easy axis and this polarization in the pinned layer biasing configuration, we shall write this spin polarization as $\Pi\mathbf{p_x} + \Pi\mathbf{p_y}$, where $\mathbf{p_x}$ and $\mathbf{p_y}$ are vectors along *(x)* and *(y)*, with their norm obeying $p_x^2+p_y^2=1$. We use a sinusoidal angular dependance of the spin transfer torque, and the standard Landau-Lifshitz-Gilbert equation:

$$\text{Eq. 1} \quad \frac{d\mathbf{m}}{dt}=[(\frac{\mu_B}{tM_S})(\frac{J_{applied}\Pi}{|e|})(\mathbf{p_y}\times\mathbf{m}+\mathbf{p_x}\times\mathbf{m})+\gamma_0\mathbf{H_{eff}}]\times\mathbf{m}$$

where $\mu_B$ is the Bohr magneton, and $\gamma_0$ is the gyromagnetic ratio, here chosen positive. We write the dimensionless current as:

$$\text{Eq. 2} \quad \tilde{J}=\frac{\mu_B}{tM_S}\frac{J_{applied}\Pi}{|e|}\frac{1}{\gamma_0 M_S}$$

Recalling that the zero temperature instability current in the standard configuration is $J_{C0}\approx\alpha(\mu_0 M_S^2 t|e|)/(2\Pi\hbar)=10^{11} A/m^2$, we have typically

$$\text{Eq. 3} \quad \tilde{J}=\frac{\alpha}{4}\frac{J_{applied}}{J_{C0}}=\frac{\alpha(\Delta+1)}{4}$$

where $\Delta=(J_{applied}-J_{C0})/J_{C0}>0$ is the overdrive, of the order of a few times unity for sub-nanosecond switching situations. The order of magnitude of $\tilde{J}$ will thus be typically that of the damping parameter $\alpha$.
Using these notations and a dimensionless time $\tau=\gamma_0 M_S t$, the projections of the dynamical equations reduce to:

$$\text{Eq. 4} \quad \dot{m}_x=\tilde{J}(-p_x m_z^2-p_x m_y^2+p_y m_x m_y)+m_z m_y+h_y m_z$$

$$\text{Eq. 5} \quad \dot{m}_y=\tilde{J}(p_x m_x m_y-p_y m_x^2-p_y m_z^2)-(1+h_k)m_z m_x$$

$$\text{Eq. 6} \quad \dot{m}_z=\tilde{J}(p_y m_y m_z+p_x m_x m_z)+h_k m_x m_y-h_y m_x$$

where the dot superscript denotes the derivative with respect to the dimensionless time $\tau$.

### 3. Switching durations versus initial magnetization orientation in the standard configuration

Before discussing the complex situations when either the free layer or the pinned layer is biased, we would like to remind the main characters of the switching in the standard configuration, i.e. when $p_x=1$, $p_y=0$ and $H_y=0$.

Figure 2A displays the calculated switching times in the standard configuration for a current $J_{applied}$= -3×10$^{11}$ A/cm² applied on 10$^4$ initial configurations **m$_0$** ={$m_{x0}$, $m_{y0}$, $m_{z0}$} taken within the interval $m_{y0}$ ∈ [-0.5, 0.5] and $m_{z0}$ ∈ [-0.05, 0.05]. The overall switching duration map looks like a dual-spiral galaxy. It contains spiraling stripes in which the switching duration evolves continuously (locally uniform grey level). It also contains spiraling contours where the switching duration evolves discontinuously (steps of grey level). This spiraling pattern reflects the underlying switching trajectories: as a response to the current step, the magnetization performs a precessional motion around the easy axis, and the gradually growing amplitude of the precession results in a spiral-like trajectory [11] when projected in the {$m_y$, $m_z$} plane. The required Number of Precession Cycles for switching (NPC) generally increases or decreases continuously when the initial state {$m_{y0}$, $m_{z0}$} is slightly varied, but the NPC is systematically incremented or decremented by ½ when the initial state {$m_{y0}$, $m_{z0}$} is moved across a spiraling contour of the switching duration map. The two spirals converge to a common Spiral Center **m$_{SC}$**, which is {1, 0, 0} in the standard configuration.

Both the magnetization position **m$_{SC}$** and the ones along the spiral contours are interesting because they lead to peculiar magnetization trajectories passing through a position where the total torque acting on the magnetization vanishes. In the case of an initial magnetization position at the spiral center, the initial torque vanishes, while in the case of an initial magnetization position along a spiral contour, the initial torque is finite but the torque acting on the magnetization at the switching instant vanishes. Both types of initial states lead to a divergence of the switching time.

(i) When the total torque cancellation happens at the pulse onset (**m$_0$**=**m$_{SC}$**), not only the switching time diverges but so does the NPC.

(ii) Conversely when an initial magnetization is along a spiral contour, the total torque cancellation happens at the switching instant [9], and the switching time diverges but the NPC remains finite and dual-valued. Let us recall the explanation of this counter-intuitive fact. When the hard axis is overcome, the magnetization is slightly out of the plane and we have typically $m_y \approx \pm 1$. If the $m_z$ component of the magnetization at the switching instant is well chosen, the demagnetizing and the spin torques can cancel each other; the magnetization feels then a vanishing total torque, and a perturbation is needed to either switch immediately or perform another half precession cycle before indeed switching. In a former publication [9] the current required for a given initial {$m_{y0}$, $m_{z0}$} to be along a spiraling contour was referred as the current leading to *maximum jitter* in the switching duration, because initial conditions obeying this criteria do lead to NPC defined with a ½ unit uncertainty. Despite the fact that the switching time diverges, the NPC remains finite along the spiral contours.

We shall see that this concept of Spiral Centers and Spiral Contours where the switching duration diverges is quite general, and that it holds in particular for the two biased configurations considered in the present work.

## 4. Implementation of the thermal fluctuations

In this section, we describe how we deal with the finite temperature, and we detail the consequence of temperature on the statistical distribution of the switching duration in the standard configuration.

### *4.1. Statistical distribution of initial magnetizations*

Before the current step is applied to the system, we assume that the initial magnetization **m$_0$** ={$m_{x0}$, $m_{y0}$, $m_{z0}$} is not far from one of its equilibrium positions, i.e.

$\mathbf{m^{EQ}} = \{m_x^{EQ}>0, m_y^{EQ}=h_y/h_k, 0\}$. With the material parameters used in this study, which are similar to that of many experiments, the thermal stability factor $1/2\, H_k M_S V / k_B T$ is only 50, such that the magnetization vector significantly wiggles and fluctuates around its equilibrium position.

Let us write the standard deviation of the magnetization to its equilibrium position as $\{\Delta m_y^{TH}, \Delta m_z^{TH}\}$. These values can be derived [12] from energy equipartition, letting both the anisotropy energy and the demagnetization energy fluctuate by $k_B T/2$. The free layer magnetization is thermally distributed inside a cone of typical aperture $\Delta m_z^{TH}=\sqrt{(k_B T)/(\mu_0 M_S^2 V)}\approx 0.01$ and $\Delta m_y^{TH}=\sqrt{(k_B T)/(\mu_0 H_K M_S V)}\approx 0.1$ (numerical values correspond to $T$=300K). The aperture of this cone is sketched as the green ellipse in Fig. 2A.

In our calculations of the statistical distribution of switching times, we consider a Boltzmann distribution of initial magnetization states regularly meshed in the $\{m_{y0}, m_{z0}\}$ parameter plane around the equilibrium magnetization position $\mathbf{m_{EQ}}$. We will assume that the probability of encountering an initial state within the volume $[m_{y0}, dm_{y0}]\times[m_{z0}, dm_{z0}]$ scales with $e^{\frac{-E}{k_B T}} dm_{y0} dm_{z0}$ where $E$ is the total energy of the system:

$$\text{Eq. 7} \quad E=\mu_0 M_S^2 V \left(-h_y m_{y0} + \frac{1}{2} h_k m_{y0}^2 + \frac{1}{2} m_{z0}^2\right)$$

For each initial state, we take the switching time (Fig. 2A) as formerly calculated from the deterministic dynamical equations Eq. 4 to Eq. 6. We then assign each obtained switching time a probability according to the statistical occupancy of the initial state, and construct an histogram of the weighted switching times. Integration of this weighted histogram provides the fraction of already switched events versus time for the initially chosen current density (Figure 2B).

Note that this calculation procedure means that we neglect the fluctuations thermally excited *during* the reversal, i.e. that we consider the adiabatic limit after the pulse onset. Our assumption can only be valid if the reversal is faster that the time required for coupling the macrospin dynamics to the non magnetic degrees of freedom, i.e. if the reversal is faster that both a few times the Néel-Brown attempt time $\tau_0$ and a few times the spin wave lifetime $1/\alpha\gamma_0 M_S$. In practice, our computational method can not be used reliably for switching times longer than typically 2 ns.

### *4.2. Statistical distribution of switching duration in the standard configuration*

In Figure 2B, we report the statistical distribution of switching durations for various applied current densities in the standard configuration at $T$=300K. For each applied current, the switched fraction increases steadily versus time [13]. The fastest switching events (bottom end of the curves in Figure 2B) correspond to the initial states that had a non-zero probability of occupancy and were the most far away from the $\mathbf{m_{SC}}$ in Figure 2A. An important feature of the standard configuration is that since the switching time diverges for initial states either along the spiral contours or at the spiral center $\mathbf{m_{SC}}$, the switched fraction can theoretically never reach 1, however arbitrary long the current is applied. The distribution of switching durations is intrinsically unbounded in the standard configuration.

This impossibility to get perfect (100%) switching reliability will exist whatever the temperature, because it arises from the fact that the thermal distribution of initial states is centered around $\mathbf{m_{EQ}}$, which in the case of the standard configuration is similar to $\mathbf{m_{SC}}$. Unfortunately, the position of the spiral center $\mathbf{m_{SC}}$ is independent of the current in the standard configuration; thus, this impossibility to get perfect switching reliability can not be lifted by the application of any stronger current. This recalls experimental results published in the literature [7] where the distribution of switching times exhibited long tails before seeming to reach full reliability in a finite (~1000) number of tests.

Let us now study the biased configurations, in order to see whether we can solve this reliability issue.

## 5. Magnetization switching in the biased configurations

We start by looking at the main features of the switching duration in both biased configurations. We then discuss the initial states which lead to a divergence of switching time, i.e. the positions of the spiral centers. At the end of this section, we finally calculate how much current is required (or equivalently how strong the bias has to be) to expel the spiral centers outside of the distribution of initial states, with the aim of getting a distribution of switching durations that is more bounded.

### *5.1.  Switching durations versus initial state in the biased configuration*

The switching durations are mapped versus the possible initial magnetization orientations for various currents in Figure 3. Panels A to D (E to H) gather the switching durations for the PLBC (FLBC respectively). In these calculations, the biasing are supposed to be done with the same 24 degrees tilting angle for the pinned layer in the PLBC ($p_y$=0.4) and for the free layer in the FLBC ($h_y/h_k$=0.4). Similar currents are used for panels that are lined up in the same column. Aside from the global similarity of the spiraling dependences of the switching durations among all biased or unbiased configurations, several differences with the standard configuration are worth noticing.

(i) First for each biasing configuration and each applied current, there is a sole specific initial condition {$m_{y0}$, $m_{z0}$} for which the resulting switching duration diverges. As it was formerly the case in the standard configuration, this specific initial condition is at the Center of a Spiral network, and will therefore be still quoted hereafter as $\mathbf{m}_{SC}$. Note that we have displayed the magnetization equilibrium orientation $\mathbf{m}_{EQ}$ as the center of each panel of Figure 3. We emphasize that the $\mathbf{m}_{SC}$ is driven away from $\mathbf{m}_{EQ}$ as the current is increased (from left to right panels in Figure 3).

(ii) A thorough comparison of the FLBC and PLBC for the same currents (e.g. Panels A-E to D-H) indicates that the Spiral Center $\mathbf{m}_{SC}$ in the FLB configuration seems to be the symmetric of the Spiral Center $\mathbf{m}_{SC}$ in the PLB configuration with respect to the magnetization equilibrium orientation $\mathbf{m}_{EQ}$. We shall see in the next sections that this symmetry is quite general when the biasing parameters are chosen with similar resulting tilting angles of either the Free Layer or the Pinned Layer (i.e. when we choose $p_y \equiv h_y/h_k$ ).

(iii) The last point to notice in Figure 3 is the *number* of spiral contours in each panel. In the FLBC at low currents (panels E and F) there is a *single* spiral contour. It can be seen that for any pair of initial states, it is possible to draw a path to connect them without ever crossing a spiral contour. In contrast, in the standard configuration (Fig. 2A), or in the PLBC (panels A to D) and in the FLBC at high current (panels G and H), there are exactly *two* spiral contours; initial magnetization orientations belong either to one spiral stripe or to the other one. This doubling of the number of spiral contours is striking when comparing PLB and FLB at low currents, i.e. when comparing panels A and E or panels B and F.

The number of spiral contours is instructive since it reflects the nature of the switching path. Indeed, in the standard and PLBC configurations, switching by passing the hard axis near $m_y$=+1 or near $m_y$=-1 can both occur depending on the initial magnetization orientation. These *two* ways of overcoming the hard axis lead to *two* spiral contours in the switching duration maps (Figure 2B and Figure 3ABCDGH). Moving an initial state across a spiraling contour makes the switching trajectory change from positive hard axis overcoming (switching near $m_y$=+1) to negative hard axis overcoming (or vice versa); such crossing of a spiraling contour increments the number of precession cycles NPC needed for switching by units of $\pm 1/2$. This no longer applies in the FLBC at low current: in that case, the reversal proceeds always by passing near the hard direction *favored* by the biasing field, i.e. near $m_y$=+1 if $H_y$>0. As a result, the crossing of a spiraling contour increments the NPC by units $\pm 1$ and there are twice less spiraling contours.

From the previous section, we could anticipate that if we aim at a satisfactory switching reliability, we need to expel the spiral center **m**$_{SC}$ away from the centre **m**$_{EQ}$ of the thermal distribution of initial states. Let us thus determine **m**$_{SC}$ in both biasing configurations.

### *5.2. Zero torque positions and spiral centers in the biased configurations*

To find the magnetization positions where the total torque vanish, resulting in a static magnetization and an infinite switching duration, we simply need to say that Eq. 4 to Eq. 6 are zero.

In the FLBC, the zero torque position **m**$_{SC}$ satisfies:

$$\text{Eq. 8} \quad m_y^{SC} = \frac{h_y(1+h_k)}{h_k(1+h_k)+\tilde{J}^2} \quad \text{and} \quad m_z^{SC} = \frac{h_y \tilde{J}}{h_k(1+h_k)+\tilde{J}^2}$$

As expected, **m**$_{SC}$ converges to **m**$_{EQ}$ in the limit of vanishing current. An interesting expression is how much spacing the current creates between **m**$_{SC}$ and **m**$_{EQ}$. These spacings are reported in Figure 4, for free layer biasing at 12 degrees ($h_y/h_k$=0.2) and at 24 degrees ($h_y/h_k$=0.4). The SC to equilibrium position $m_z$ spacing is simply Eq. 8B and scales almost linearly with the current, while the $m_y$ spacing scales almost quadratically with the current:

$$\text{Eq. 9} \quad m_y^{SC} - m_y^{EQ} = -\left(\frac{h_y}{h_k}\right)\frac{\tilde{J}^2}{h_k(1+h_k)+\tilde{J}^2}$$

In the PLBC, the zero torque position **m**$_{SC}$ can unfortunately not be expressed analytically without approximation. However if we consider that $m_x^{SC} \approx 1$ and $|m_y^{SC}| \gg |m_z^{SC}|$, then the approximate zero torque position can be described simply by:

$$\text{Eq. 10} \quad m_y^{SC} \approx \frac{\tilde{J}^2 p_x p_y}{h_k(1+h_k)+\tilde{J}^2 p_x^2} \quad \text{and} \quad m_z^{SC} \approx \frac{-\tilde{J} h_k p_y}{h_k(1+h_k)+\tilde{J}^2 p_x^2}$$

To confirm the validity of these approximate expressions, a numerical evaluation of the SC versus the applied current density has been done in Figure 4, for pinned layer biasing at 12 degrees ($p_y$=0.2) and at 24 degrees ($p_y$=0.4).

The comparison between the **m**$_{SC}$ to **m**$_{EQ}$ spacings in FLBC and PLBC is very instructive. Indeed, when the biasing of the FL and of the PL is done with the same resulting small tilting angles, i.e. by choosing $p_y=h_y/h_k \ll 1$ and $p_x \approx 1$. In that case, the vectors **m**$_{SC}$ - **m**$_{EQ}$ are opposite to each others in the FLBC and PLBC (compare Eq. 8B, and Eq. 10B and Eq. 9A and Eq. 10A). This symmetric placement of **m**$_{SC}$ with regards to **m**$_{EQ}$ in FLBC and PLBC was already anticipated in alinea (ii) of §5.1 from a close comparison of the top panels versus the bottom panels of Figure 3. It was not accidental.

### *5.3. Expelling zero torque positions outside of the thermal distribution of initial states*

To get a more bounded distribution of switching durations, we need at least to push $\mathbf{m_{SC}}$ away from the core of the thermal distribution of initial states. Ideally, we would like to apply biasing parameters and current in order to satisfy either $|m_y^{SC} - m_y^{EQ}| \gg \Delta m_y^{TH}$ or $|m_z^{SC}| \gg \Delta m_z^{TH}$ (remember $m_z^{SC}=0$). We will quote these criteria as « zero initial susceptibility avoidance » criteria. These zero initial susceptibility avoidance criteria are evaluated numerically in the two panels of Figure 4, where the double of the standard deviations of the initial magnetization components { $\Delta m_y^{TH}, \Delta m_z^{TH}$ } are displayed as the shaded areas.

In the PLBC, the first zero initial susceptibility avoidance criterion requires to apply current much higher than:

$$\text{Eq. 11} \quad |\tilde{J}_{expel\,Z}| = \frac{1}{p_y} \sqrt{\frac{\mu_0 M_S^2 V}{k_B T}}$$

The second zero initial susceptibility avoidance criterion requires to apply current much higher than:

$$\text{Eq. 12} \quad \tilde{J}_{expel\,Y}^2 = \frac{h_k}{p_y} \sqrt{\frac{\mu_0 M_S H_k V}{k_B T}}$$

In the FLBC, similar zero initial susceptibility avoidance criteria can be obtained by replacing $p_y$ by $h_y/h_k$ in the previous two equations.

With $p_y = h_y/h_k = 0.4$, numerical evaluations of the criteria for avoidance of zero initial susceptibility yield $|\tilde{J}_{expel\,Z}| = 0.025$ and $|\tilde{J}_{expel\,Z}| = 0.05$. Only the smallest values of the two latter currents is required to expel the spiral center from the core of the thermal distribution. Using Eq. 2, we find that to push the spiral center $\mathbf{m_{SC}}$ at a distance greater than $\Delta m_z^{TH}$ from the $\mathbf{m_{EQ}}$, we need to apply an overdrive of $\Delta=4$. Hence, either FLBC or PLBC can be used to avoid getting frequently initial states leading to very slow switching events. However, even if biasing makes those unwanted events much less frequent, they still can occur with a non negligible probability. Let us thus now evaluate quantitatively the statistical distribution of switching times.

### **6. Statistical distribution of switching times in the biased configurations**

In this section, we detail the statistical distributions of switching durations in the FLB and PLB configurations. We first list the main aspects of these distributions, then we focus on the tail of the distribution at long switching times to derive predictions linking a given targeted switching error rate with the required current pulse duration. We finally compare the standard deviations of switching times to identify the optimal biasing parameters and overdrive current.

### *6.1. Description of the statistical distribution of switching durations*

The statistical distributions of switching durations are displayed in Figure 5 for various currents and various biasing strategies. Panels A and B (C and D) gather the results for the PLBC (FLBC respectively). In these calculations, the biasing are supposed to be done with a tilting angle of 12 degrees for panels A (PLB $p_y$=0.2) and C (FLB $h_y/h_k$=0.2), and a tilting angle of 24 degrees for panels B (PLB $p_y$=0.4) and D (FLB $h_y/h_k$=0.4). In Figure 6, we have drawn a direct comparison of the benefits of the PLBC and FLBC strategies in terms of switching time reproducibility. From Figure 5 and Figure 6, we could draw several conclusions:

*(i)* Our first conclusion concerns the technological interest of biasing, both in terms of average switching speed and in terms of switching reliability. It is clear from Figure 6 that the switching speed is enhanced by the biasing, as was already identified experimentally [8]. However, the main benefit of biasing is that the time required for error-free switching is strongly reduced compared to the standard configuration: the tail of the switching duration distribution at long durations is shrunk to a great extent by both biasing strategies. This reduction of the switching-error-free current pulse duration was also seen in experiments [8] and will be quantified analytically in the next sub-section.

*(ii)* Another striking aspect of the statistical distribution of switching durations is that while in the standard configuration, the switched fraction increased steadily versus time (Figure 2B), the switched fraction in the biased configurations increases stepwise, with a step-to-plateau contrast that develops as the biasing parameter $h_y/h_k$ or $p_y$ is increased, recalling former experimental results [9]. These steps simply recall that the NPC required for switching has a comb-like distribution, that is peaked at discrete time intervals.

*(iii)* At low and moderate current (-3 to -5×10$^{11}$ A/cm² ) the steps are typically both *twice* higher and *half* less numerous in the FLBC case than in the PLBC case (compare e. g. panels B and D, black curves). This is no longer true for larger currents. This is related to the number of spirals in the switching durations maps (Figure 4): at low and moderate currents in the FLBC case, the NHP values increment by units of $\pm 1$ when crossing a spiral, and this leads to plateaus that last *one mean* precession period in the statistical distribution of switching times. In all other configurations (PLBC, standard configuration, or FLBC at higher currents), the NHP values increment by units of $\pm 1/2$ when crossing a spiral, and this leads to plateaus that lasts *half a mean* precession period in the statistical distribution of switching times.

### *6.2. Pulse duration required for a targeted write error rate*

Let us now quantify the benefit of the biasing configurations in terms of switching error rate. For this we need to focus on the tail of the switching duration distribution to find out how much margin in the pulse duration should be taken to ensure almost sure switching. Recalling that the tail at long durations in the distribution of switching times is related to those initial magnetization orientations being very near the spiral center, and recalling that the spiral centers are symmetric with respect to the equilibrium magnetization orientations in the two biased configurations (Eq. 9A and Eq. 10A), one may anticipate that the tails of the statistical distributions of switching time should be very similar in the two biasing configurations. If we consider the shape of this tail as a figure of merit, a thorough comparison of the two biasing strategies (Figure 6) can not decide whether FLBC or PLBC is best; the corresponding long duration tails in the statistical distribution of switching times almost super-impose (see Figure 6A at t>0.7 ns or Figure 6B at t>0.5 ns) withing the calculation precision (probability increments of $10^{-4}$); an analytical study is thus required to understand the asymptotic shape of the tail in the statistical distribution of switching duration.

For this purpose we write $t_N$ the pulse duration needed to achieve a write error rate of $10^{-N}$, *i.e.* a switching probability of $1-10^{-N}$, with $N \geq 2$. To estimate $t_N$ we first need to list the initial magnetization positions that are sufficiently near the SC so that their switching time exceeds $t_N$. Near a vanishing total torque position like $\mathbf{m_{SC}}$, there is unfortunately no general expression describing how the switching time diverges. However, previous studies [11] indicated that the general trend is a divergence of the switching time like $\frac{1}{\alpha \gamma_0 M_S \Delta} \log \frac{\Xi}{\|\mathbf{m}-\mathbf{m}_{SC}\|}$ where $\Xi$ is a constant that we will not need to determine. Hence, the area of initial positions within the $\{m_{y0}, m_{z0}\}$ plane leading to reversal events lasting longer than $t_N$ scales with $\exp(-2 t_N \alpha \gamma_0 M_S \Delta)$.

The corresponding Boltzmann probability that the initial magnetization orientation be in that area is proportional to $\iint e^{-(E-E_{EQ})/kT} dm_{y0} dm_{z0}$. Since this area is small for $N \geq 2$, we can approximate the integrand by simply $e^{-(E-E_{EQ})/kT} \exp(-2 t_N \alpha \gamma_0 M_S \Delta)$. We finally can say that the pulse duration $t_N$ needed to achieve an error rate of $10^{-N}$ satisfies $10^{-N} = \Sigma \exp(-(E_{SC}-E_{EQ})/kT) \exp(-2 t_N \alpha \gamma_0 M_S \Delta)$ where $\Sigma$ is a constant.

From this former expression, we can see that in order to achieve the lowest switching error rate, we have to maximize the pulse duration $t_N$, the overdrive current $\Delta$, and the *energy* spacing $E_{SC}-E_{EQ}$ between the magnetization orientation at equilibrium and at the spiral center SC. Maximizing this energy spacing can be done by either increasing the biasing parameter ($h_y/h_k$ of $p_y$) or the overdrive current $\Delta$. Note that a close look [14] at the energy functional (Eq. 7) indicates that it is equivalent to maximize the *energy* spacing $E_{SC}-E_{EQ}$ and the *magnetization* spacing $\|\mathbf{m_{SC}}-\mathbf{m_{EQ}}\|$, as was done in §5.3 to derive the zero initial susceptibility avoidance criterion (Eq. 11 and Eq. 12). To calculate the relation between switching error rate and current pulse duration, we only need to know any [15] of the $t_N$. For instance, from Figure 6A we can read $t_2$=793 ps for 99% switching probability in the PLBC with $p_y$=0.4 and $J_{applied}$=6 ×$10^{11}$ A/m². The relation between switching error rate $10^{-N}$ and the required current pulse duration $t_N$ is thus:

$$\text{Eq. 13} \quad t_N \approx t_2 + (N-2) \frac{\log(10)}{2 \alpha \gamma_0 M_S \Delta}$$

Note that in the above expression $t_2$ is a function of $\Delta$ and of the biasing parameters, hence it must be determined from the procedures described in section §4.2.

With the material parameters of this study ($1/\alpha \gamma_0 M_S = 0.28 \, ns$) a targeted error rate of $10^{-9}$ in the PLB configuration with $p_y$=0.4 and $J_{applied}$=-6 ×$10^{11}$ A/m² requires a pulse duration of 1.25 ns.

As stated previously, since we consider the adiabatic limit, our predictions can not be quantitative when dealing with switching times longer than a few times $1/\alpha \gamma_0 M_S$. Since Eq. 13 does not take into account thermal fluctuations after the pulse onset, it will underestimate the switching probability at long duration times. Eq. 13 should thus be regarded as a *sufficient* condition to achieve the targeted error rate but should not be considered as a *necessary* condition.

### 6.3. *Optimal choice of biasing parameters to minimize the standard deviation of the switching time*

Our last comment concerns the relevant choice of current and biasing parameters to minimize the standard deviation of the switching duration. To achieve this goal, we need to avoid getting switching events with differing NPC in the thermal distribution of initial states. Geometrically, this criteria is simply a « spiraling contour avoidance » criterion, expressing that we require avoiding to have any spiral contour intersecting the thermal distribution of initial state.

Let us show situations and see whether they satisfy of this criterion.

The worst (greatest) standard deviations of switching duration are obtained when a spiral contour passes exactly through or very near $\mathbf{m_{EQ}}$. This is for instance the case for PLB configuration and $J_{applied}$=-6×10$^{11}$ A/m² (Figure 4C): the switching times apart the spiral contour near $\mathbf{m_{EQ}}$ are 300 ps and 450 ps. These two values can immediately be noticed as clear steps in the corresponding statistical distribution of switching times (Figure 6A).

The best (smallest) standard deviations of switching duration are obtained when all spiral contours pass quite far from $\mathbf{m_{EQ}}$. This is for instance the case for the FLBC at $J_{applied}$ = -6 ×10$^{11}$ A/cm² (Figure 4G). The corresponding distribution of switching times is the narrowest among our studied parameters, and is much better than in the standard configuration (Figure 6A). Before concluding, we wish to emphasize that the spiraling contour avoidance criterion is not easy to implement, since it requires a fine trade-off between the biasing and the current. For instance, we have increased the current up to $J_{applied}$=-8×10$^{11}$ A/cm² in Figure 6B, and despite the fact that the zero initial susceptibility avoidance criterion is better satisfied, the spiraling contour avoidance criterion is not satisfied any longer, and the statistical distribution of switching durations show steps and plateaus for both biased configurations.

### 7. **Summary and concluding remarks**

In this paper, we have studied the statistical distribution of the switching durations in spin-transfer switching. We have compared three model spin-valves: a spin-valve with a free layer whose easy axis is parallel to the pinned layer magnetization (standard configuration), a pinned layer with magnetization tilted with respect to the free layer easy axis (pinned layer biasing configuration PLBC), and a free layer whose magnetization is pulled away from easy axis by a hard axis bias (free layer biasing configuration FLBC).

We have shown that in each of these three configurations, there is an initial magnetization position $\mathbf{m_{SC}}$ for which the initial total torque acting on the magnetization vanishes (Eq. 8 and Eq. 10) leading a vanishing initial susceptibility when the spin torque is applied. Within the macrospin and adiabatic approximations, the switching duration diverges if the initial magnetization is chosen along that orientation.

In the standard configuration $\mathbf{m_{SC}}$ matches with the equilibrium position of the magnetization $\mathbf{m_{EQ}}$. At finite temperature, the initial magnetization position is distributed around $\mathbf{m_{EQ}}$, such that the distribution of switching durations includes a switching event with an infinite switching time. Therefore, the distribution of switching durations can not be made narrow, however strong the current amplitude be.

In the PLBC and the FLBC, the statistical occurrence of initial magnetization states $\mathbf{m_{SC}}$ with zero susceptibility to the spin torque can be minimized, provided that the biasing parameters and the current are chosen strong enough (and Eq. 11 and Eq. 12) that the $\mathbf{m_{SC}}$ positions be as far as possible from the core of the thermal distribution of initial magnetization states. When these zero initial susceptibility avoidance criteria are satisfied, the tails of the statistical distribution of switching durations are much reduced, which is a clear benefit when aiming at error-free switching. A conservative estimate of the required pulse duration needed to achieve a target in switching error rate has been derived (Eq. 13).

      Finally, we have studied how the choice of the current and of the biasing parameters affects the standard deviation of the switching duration. We have emphasized that the number of precession cycles before overcoming the hard axis, hence the time required for switching strongly depends on the initial magnetization state. This can be seen as spiral contours in the dependence of the switching time versus the initial magnetization (Figure 4). Since the initial magnetization states is distributed according to the finite temperature, this can lead to a multiply-stepped statistical distribution of switching time, recalling earlier experimental results [9]. The multi-step behavior is most pronounced when the magnetization equilibrium position leads to an incremental jitter in the switching time, i.e. when the magnetization equilibrium position matches with a spiral contour. Using a subtle trade-off between the current amplitude and the biasing strength in the FLBC, we can almost satisfy the criterion of spiraling contour avoidance. This is realized in the free layer biasing configuration at high applied current density. The lowest standard deviation of switching duration is then achieved, showing the clear benefit of the free layer biasing strategy.

**Figures**

## Figure 1

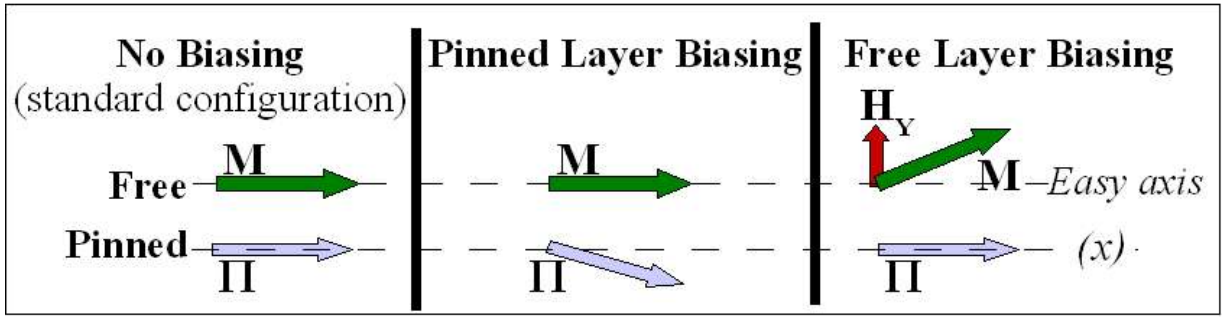

Figure 1: Sketch of the biasing configurations. Left panel: standard configuration with no applied field, and free layer magnetization **M** along its anisotropy axis *(x)*, chosen parallel to the pinned layer spin polarization Π. . Center panel: pinned layer biasing (PLB) configuration. Right panel: free layer biasing configuration, where a biasing field (either exchange field or external field) is applied along *(y)* the direction of the hard axis of the free layer.

___________________________________________

# Figure 2

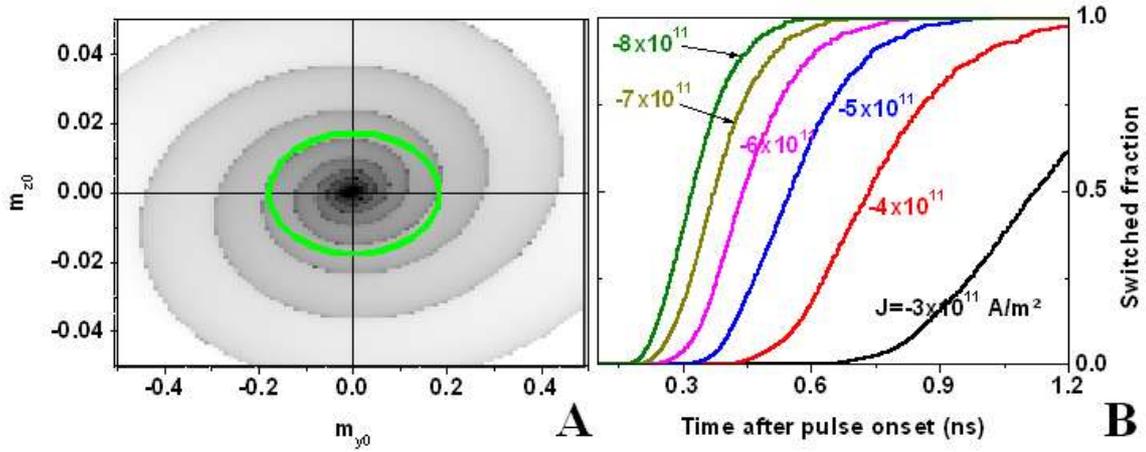

Figure 2: **A** : Switching duration versus initial magnetization state in the standard configuration for $J_{applied}$=-3×10$^{11}$ A/m². The gray level is scaled to white for 400ps or less switching duration, and to black for 2 ns or more switching duration. Considering thermal fluctuations, most initial states will be inside the core of the thermal distribution that is sketched as the green ellipse superimposed on the switching duration map. **B** : Probability of being already switched at a given variable time after the onset of the pulsed current. Calculation is done for the standard configuration when the applied current is $J_{applied}$=-3×10$^{11}$ A/m² at T=300 K.

Figure 3

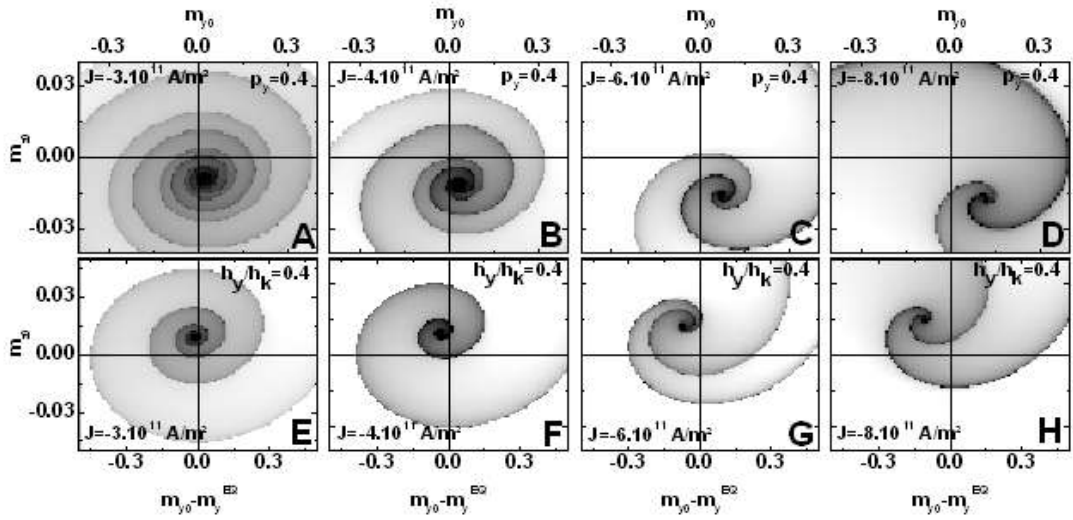

Figure 3: Switching durations versus initial magnetization state in the pinned layer biasing configuration (panels **A** to **D**) and in the free layer biasing configuration (panels **E** to **H**). The the applied current increases from left panels to right panels. The gray level is scaled to white for 400 ps switching (panels **A**, **B**, **E** and **F**), for 300 ps (panels **C** and **G**) and for 150 ps (panels **D** and **H**). The gray level is scaled to black for 2 ns (panels **A** and **B**), for 1.2 ns (panels **B** and **F**), for 800 ps (panels **C** and **G**) and for 600 ps (panels **D** and **H**). The centre of each panel corresponds to the equilibrium position of the magnetization.

# Figure 4

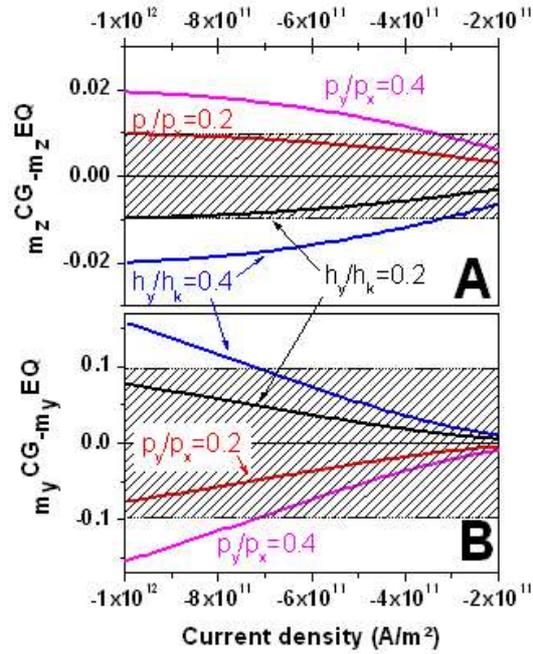

Figure 4: Projections along $m_z$ (panel **A**) and along $m_y$ (panel **B**) of the distance between the spiral center $\mathbf{m}_{SC}$ and the equilibrium magnetization position $\mathbf{m}_{EQ}$ for the two biasing configurations versus the applied current density. The shaded areas have widths that are twice the standard deviation $\Delta \mathbf{m}_{TH}$ of the initial magnetization away from equilibrium position.

# Figure 5

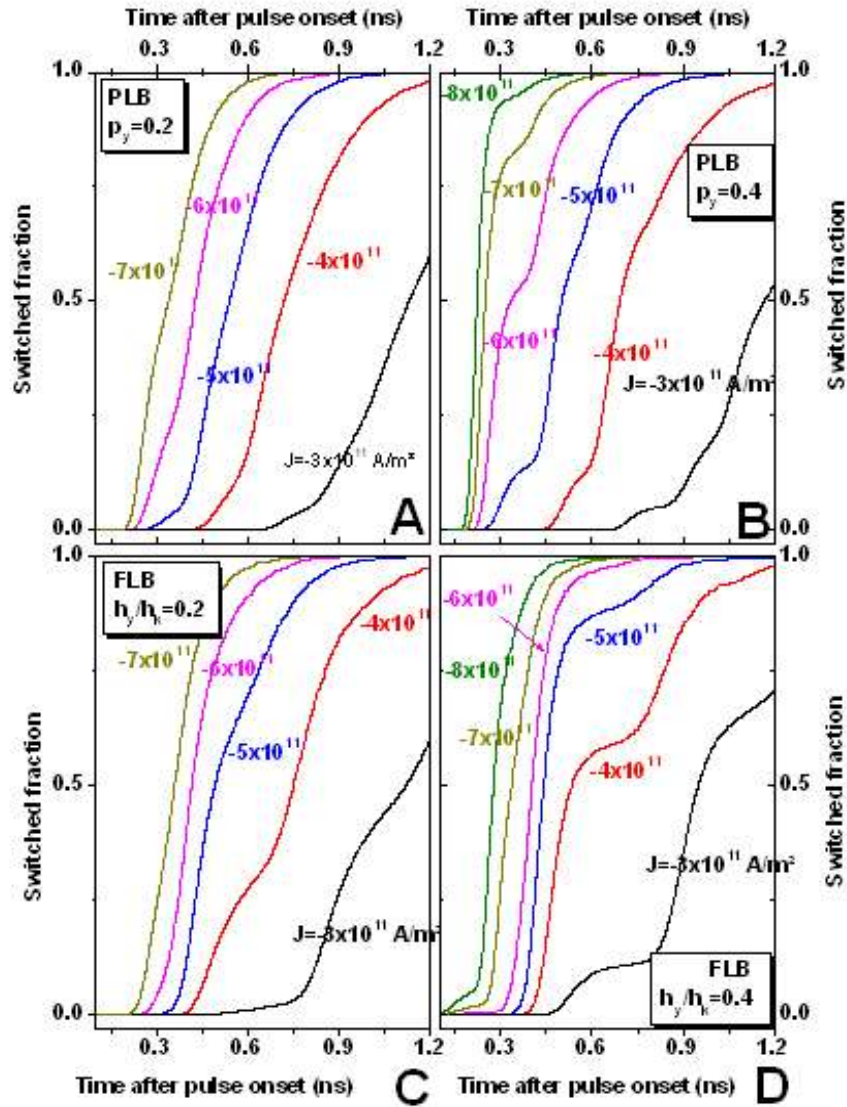

Figure 5: Statistical distributions of the switching duration versus applied current for the pinned layer biasing configuration with $p_y$=0.2 (panel **A**) or with $p_y$=0.4 (panel **B**), and for the free layer biasing configuration with $h_y/h_k$=0.2 (panel **C**) or with $h_y/h_k$=0.4 (panel **D**). The calculations are done at T=300 K.

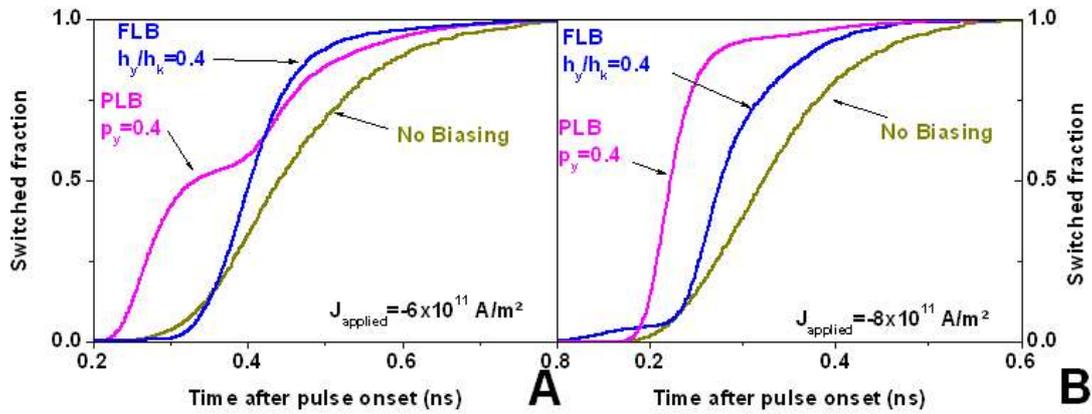

Figure 6: Comparison of the statistical distribution of the switching duration in the standard configuration, free layer biasing configuration and pinned free biasing configuration at T=300 K for $J_{applied}$=-6×10$^{11}$ A/m² (panel **A**) and for $J_{applied}$=-8×10$^{11}$ A/m² (panel **B**).